\documentclass{cimento}

\usepackage{graphicx}  

\title{Prospects for an experiment to measure $BR(K_L\rightarrow \pi^0\nu\bar\nu)$ at the CERN SpS}
\author{S.~Martellotti\from{ins:x} 
on behalf of the NA62 Collaboration.}
\instlist{\inst{ins:x} INFN, Laboratori Nazionali di Frascati - Frascati, Italy.}

\begin{document}

\maketitle

\begin{abstract}
We are investigating the feasibility of performing a measurement of $BR(K_L\rightarrow\pi^0\nu\bar\nu)$ 
using a high-energy secondary neutral beam at the CERN SPS in a successor experiment
to NA62. 
The timescale would require many years; 
we assume that the experiment would be ready at
the start of LHC Run 4. Some preliminary conclusions from our feasibility studies, design challenges faced and sensitivity obtainable for the measurement are here presented. 
\end{abstract}

\section{Introduction}
Precise measurements of the branching ratios for the $K\rightarrow \pi\nu\bar\nu$ decays can provide
unique constraints on CKM unitarity and, potentially, evidence for new physics. It is important
to measure both decay modes, $K^+\rightarrow \pi^+\nu\bar\nu$ and $K^0_L\rightarrow\pi^0\nu\bar\nu$, since different new physics
models affect the rates for each channel differently. 
NA62 experiment at the CERN SPS is now running with the goal of measure the $BR(K^+\rightarrow\pi^+\nu\bar\nu)$. In the same site, a new experiment using a high-energy secondary neutral beam could be performed for the measurement of $BR(K^0_L\rightarrow\pi^0\nu\bar\nu)$ with a 
technique complementary to the upgrade of KOTO experiment at J-PARC \cite{ref:koto} and a comparable sensitivity.

\section{Plans for a measurement of $BR(K_L\rightarrow\pi^0\nu\bar\nu)$ at SPS}

\begin{figure}[htbp]
\begin{minipage}{0.73\textwidth} 
\includegraphics[width=9.2cm]{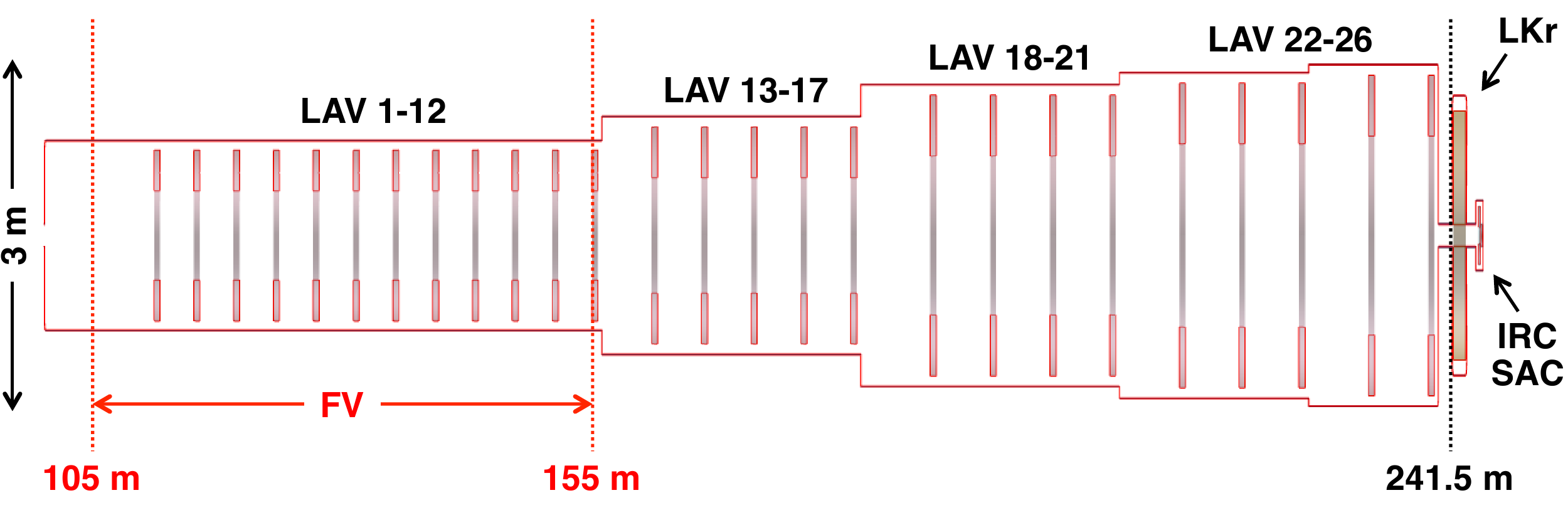} 
\caption{Schematic layout of the experiment.} 
\label{fig:setup} 
\end{minipage} 
\hfil
\begin{minipage}{0.25\textwidth} 
\includegraphics[width=3.0cm]{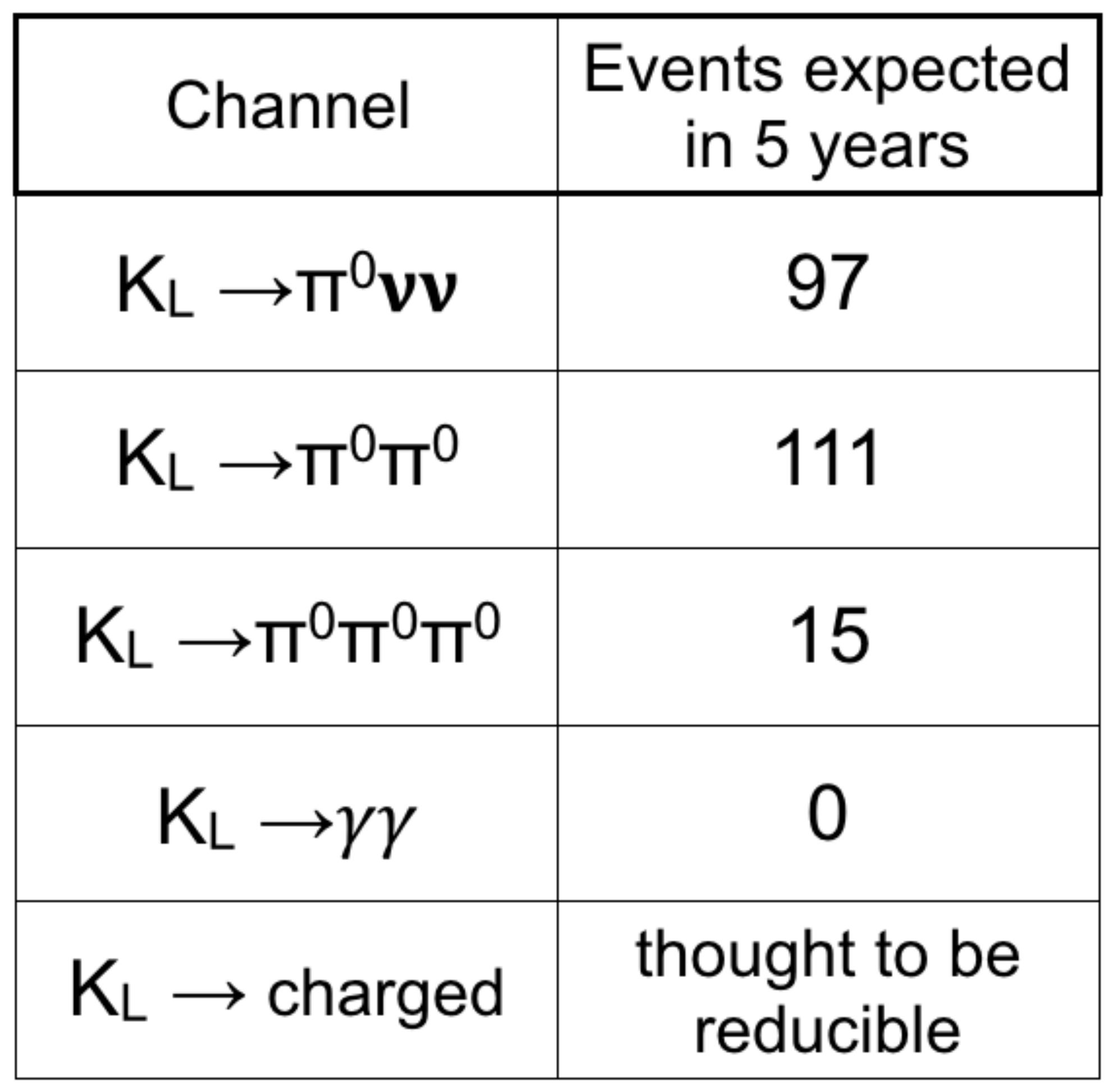}
\caption{Estimated final selected events.} 
\label{fig:tab}
\end{minipage} 
\end{figure}

As in NA62, the proposed experiment makes use of a
400 GeV primary proton beam interacting on a 400mm beryllium rod target. It impinges
the target with an angle of 2.4 mrad, which optimizes the $K_L$ to neutron and photon flux
ratios according to parameterization in \cite{ref:ath} and FLUKA simulation. The beam line has
been designed to have a system of three collimators and an absorber to reduce the huge
rate of photons. The secondary beam polar angle acceptance is 0.3 mrad and contains
about $2.8 \times 10^{−5}\, K_L$s per proton incident on the target (pot). The $K_L$ momentum
distribution peaks around 35 GeV/c and has a mean at about 97 GeV/c. There will be
$6.3\times 10^{−7}\, K_L$ decays in the fiducial volume per pot. Assuming the SM value
of $BR(K_L\rightarrow\pi^0\nu\bar\nu)$ and an acceptance fraction of 10\%, to observe 100 signal events, about $3 \times 10^{13}\, K_L$s must decay in the fiducial volume. Thus, an integrated
proton flux of $5\times10^{19}$ pot is required, which we assume is delivered at a rate of $10^{19}$ pot/yr
over the course of five years. This intensity is 6 times larger than the current NA62 one,
therefore an extensive upgrades to the beam line cavern will be needed. The planned experiment would reuse the NA48 liquid-krypton calorimeter and will be composed by a system of photon veto detectors,
a sketch is shown in Fig.\ref{fig:setup}. The $K_L$ momentum is broadly
distributed, and a  large fraction of background photons are emitted at large polar
angle. The photon veto system is required to cover a polar angle up to 100 mrad and
to be highly efficient down to 100 MeV. These requirements are not satisfied by the
NA62 Large Angle Vetoes (LAVs), and we plan to construct 26 new LAVs with a different geometry.
One possible design for the LAVs would be similar to the Vaccum Veto System detectors planned for the CKM experiment at Fermilab \cite{ref:fermilab}. 
Main challenge of the experiment will be the design of detector to veto the photons
at small polar angle (SAC), where the beam neutron and photon fluxes are very high. The veto
is needed down to 0.4 mrad and must intercept also photons from $K_L$ passing through
the beam pipe. According to a preliminary FLUKA and GEANT4 simulation of the
beam line, the beam contains 3 GHz of neutrons and 700MHz of photons to which the
detector should be insensitive. The design studies for such veto detector are ongoing.
Furthermore an Intermediate Ring Calorimeter (IRC) should not intercept the beam but
should cover the LKr bore to detect photons from downstream decays. A fast simulation
study has been performed in order to estimate the most important background from $K_L$
decays, which turns out to be $K_L \rightarrow\pi^0\pi^0$. Events are selected if there are exactly two hits in the LKr and no hits on any of the veto detectors. By imposing the $\pi^0$ mass, the invariant mass of
the two photons is used to get the z position of the $\pi^0$ vertex, that is required
to be in the fiducial volume.
After all the kinematic cuts, out of $9.6 \times 10^8$ generated signal events, $1.9 \times 10^6$ are selected, while out of $1.2 \times 10^{12}$ generated $K_L \rightarrow\pi^0\pi^0$, 85 are selected. 
The expected number of selected events in five years of data taking 
for the signal (SM BR assumption) and the main background modes are listed in the table in
Fig.\ref{fig:tab}.

Further study and a better optimization are still needed, but from these preliminary studies
we can conclude that a measurement of $BR(K_L\rightarrow\pi^0\nu\bar\nu)$ at CERN SPS is feasible with no major change in the NA62 infrastructure.

\end{document}